\documentclass[twocolumn,10pt,aps,preprintnumbers,showpacs,pre,floatfix,amsmath,amssymb,notitlepage,superscriptaddress]{revtex4-1}
\usepackage{epsfig}
\usepackage{subfigure}
\usepackage{amsmath}
\usepackage{color}
\usepackage{amssymb}
\usepackage{setspace}
\usepackage{graphicx}
\usepackage{dcolumn}
\usepackage{bm}
\usepackage{times}
\usepackage{enumerate}
\usepackage{footnote}
\usepackage{cancel}
\usepackage{epstopdf}
\input epsf

\newcommand{\appropto}{\mathrel{\vcenter{
  \offinterlineskip\halign{\hfil$##$\cr
    \propto\cr\noalign{\kern2pt}\sim\cr\noalign{\kern-2pt}}}}}

\definecolor{maria}{rgb}{0.24, 0.71, 0.54}

\begin{document}

\preprint{APS/123-QED}

\title{On the evolution of cooperation under social pressure in multiplex networks}

\author{Mar\'{i}a Pereda}
 \email{mpereda@math.uc3m.es; http://www.gisc.es}
\affiliation{%
 GISC. Departamento de Matem\'{a}ticas, Universidad Carlos III de Madrid, 28911 Madrid, Spain\\
}%




\date{\today}

\begin{abstract}
In this work, we aim to contribute to the understanding of the human pro-social behavior by studying the influence that a particular form of social pressure ``being watched'' has on the evolution of cooperative behavior. We study how cooperation emerge in multiplex complex topologies by analyzing a particular bidirectionally-coupled dynamics on top of a two-layers multiplex network (duplex). The coupled dynamics appears between the Prisoner's Dilemma game in a network, and a threshold cascade model in the other. The threshold model is intended to abstract the behavior of a network of vigilant nodes, that impose pressure of being observed altering hence the temptation to defect of the dilemma. Cooperation or defection in the game also affects the state of a node of being vigilant.
We analyze these processes on different duplex networks structures and assess the influence of the topology, average degree and correlated multiplexity, on the outcome of cooperation. Interestingly, we find that the social pressure of vigilance may impact cooperation positively or negatively, depending on the duplex structure, specifically the degree correlations between layers is determinant. Our results give further quantitative insights in the promotion of cooperation under social pressure.
\end{abstract}

\maketitle


\section{\label{sec:intro}Introduction}
%

Human cooperation is a ubiquitous yet not fully-understood phenomenon. Explaining how cooperation emerges and withstands selfish behaviors is one of the biggest challenges in natural and social sciences.
Multiple mechanisms have been proposed to explain under which conditions cooperation emerges and is sustained: direct reciprocity (repetition), indirect reciprocity (reputation), spatial selection, multilevel (group) selection, and kin selection~\cite{Nowak:2006bt, Pennisi:2005hx,Hauert:2007js}.

Evolutionary Game Theory~\cite{Smith:1982ww,Szabo:2007uy,NowakEvolDyn} is the arena to analyze the evolution of cooperation. In the last years, the analytical results of the theory are faced against experimental studies with humans facing game theoretical dilemmas~\cite{Grujic2010, GraciaLazaro2012, suri2011cooperation, rand2011dynamic}. Interestingly enough, these experiments have challenged the way we understand human cooperation, and more work on the consequences of these experiments have to follow.

Another way to approach the understanding of the evolution of cooperation in human societies consist in deciphering the cooperative behavior in ancient communities from historical records. In a previous work~\cite{nuestroPLOS} we studied cooperation in the Yamana society that inhabited the Beagle Channel in Argentina, with respect to sharing beached whales (a scarce, unpredictable and valuable resource). In that work we observed that the emergence of an informal network of vigilance promoted cooperation. 

Historically, ancient societies have exploited the power that images of watchful eyes have on people. We can find examples in totem monuments decorated with eyes to enhance charitable behaviors in tribes~\cite{NowakSupercooperators}; in different religions using this power promote honesty~\cite{NowakSupercooperators}, which is coherent with the \textit{Supernatural Monitoring Hypothesis}, which states that the perception of being watched promotes pro-social behavior ~\cite{Rossano2007, Johnson2009}.``They remind us that our actions have consequences"~\cite{NowakSupercooperators}.

The essential idea is that being watched can play an important role in promoting pro-social cooperative behavior. Several field studies have found evidence of humans exposing a pro-social behavior when being observed by others (recently confirmed in a field experiment with 2,000 individuals~\cite{Yoeli2013}) and also under the presence of subtle cues of being watched. Although there are also some studies that could not find such evidence. A review on the topic can be found at~\cite{Jaeggi2010, Pfattheicher2015}. A possible reason for the failure of previous studies in eyes cue influence is proposed in~\cite{Pfattheicher2015}, where the authors found that people with weak public self-awareness, i.e. people not concerned about how they appear in the eyes of others, are not affected by the watching eyes phenomena. 
The observability effect (the increase of cooperation under vigilance) seems to be driven by our reputational concerns, bringing the indirect reciprocity mechanism into play. 

This work is aimed to shed light, from a complex networks perspective, on the phenomenon above mentioned, i.e. the emergence of cooperation in a networked society interacting with a network of vigilance.  The effect of the structure of interactions on different social dilemmas has been largely studied within the scope of network theory over the past years, from topology influence~\cite{jesus2007} to spatial and temporal effects~\cite{Roca:2009joa}. Recently, a new perspective for the representation of multiple types of social interactions has been proposed under the name of multiplex networks. Different kind of interactions are modeled by different interconnected layers. This approach has been successfully applied to the study of the Prisoner's Dilemma Game~\cite{GomezGardenes:2012hc} and also to the understanding of cooperation in coupled networks~\cite{wang2012evolution}.

We adopt a similar approach here, modeling our problem in the scope of multiplex network. Specifically, we investigate the interplay between two dynamical processes, an evolutionary game (a Prisoner's Dilemma) and dynamical social pressure (a vigilance network evolving according to a threshold dynamics), and the duplex structure of these interactions. 

The paper is organized as follows. We present the definition of the model in Sec. II. Results obtained by means of simulation are analyzed in Section III. Firstly, we focus our analysis on the influence of these coupled dynamics on cooperation under different monoplex network structures (subsection~\ref{sssec:mono}) and the impact that different costs of vigilance have on the outcome of cooperation (subsection~\ref{sssec:cost}). Secondly in subsection ~\ref{sssec:duplex}, the analysis moves to a duplex structure of networks, where we study how the topologies and average degrees of the different layers affect cooperation, and also how the layer-degree correlations can promote or hinder cooperation. Lastly, a modification is introduced on the vigilance dynamics, where a vigilance actor can stop being vigilant. The influence of this dynamics is studied in subsection ~\ref{sssec:giveup}. Finally we conclude with section IV by summarizing what insights are offered by our work.









%



\section{Model dynamics}
\subsection{Description of the model}

The abstracted framework for our analysis is a networked system of agents (nodes) playing a theoretical game under vigilance pressures. In particular, our agents play an evolutionary Prisoners' Dilemma (PD) game. The links define the neighborhood of the players and so to whom they are playing with. The same players involved in the game are also endowed with a state that define them as vigilant or not. The whole dynamics is composed by an interaction between the game and the spreading of the vigilant behavior. The game is divided into two phases: payoff recollection and strategy update. Each round, node $i$ can choose to play one of the two strategies, cooperation (C) or defection (D). The PD game can be defined according to its payoff matrix (entries correspond to the row player's payoffs):
\vspace{-2.5 mm}
\begin{equation}
\bordermatrix{
  & C & D \cr
C & R & S \cr
D & T & P \cr}
\label{tab:payoffMatrix}
\end{equation}
where \emph{R} represents the reward obtained by a cooperator playing against another cooperator, \emph{S} is the sucker payoff obtained by a cooperator when she plays against a defector, the temptation payoff, \emph{T}, is the payoff received by a defector when his opponent is a cooperator, and finally, \emph{P} represents the payoff obtained by a defector which engages with another defector. In the PD~\cite{Axelrod_1980a}, $T>R>P>S$. We rescale the game so that it depends on only one parameter, as it is done traditionally \cite{Nowak1992}. We define $b$ as the advantage of defectors over cooperators, being $T=b>1$. The values of \emph{R} and \emph{P} are fixed to $R=1$ and $P=0$ in order to provide a fixed scale for the game payoffs. Applying this constraint, it turns out that the selection of the remaining parameters $b$ and S enables the definition of several games according to their evolutionary stability. We will focus on the PD with S=0, then being the only parameter of the game the advantage of defectors over cooperators $b$.

For the spreading of the vigilance behavior, we will assume a cascade of imitation effect. Players activate (become vigilant) (${V}^{0 \rightarrow 1}_{i}=1$) following a Watt's threshold model \cite{Watts30042002}:
\begin{equation}\label{eq:vigilance}
{V}^{0 \rightarrow 1}_{i}({m}_{i},{k}_{i})= \left\{ \begin{array}{lcc}
             1 & if & {m}_{i}/{k}_{i} > {\theta}_{i}, \\
             \\ 0 & if &   {m}_{i}/{k}_{i} \leq {\theta}_{i}, \\
             \end{array}
   \right.
\end{equation}
where ${m}_{i}$ is the number of neighbors of the node \emph{i} that are already vigilant, ${k}_{i}$ is the degree of node \emph{i}, and ${\theta}_{i}$ the personal threshold of node \emph{i} above which she becomes vigilant. Note that we identify vigilance as a pro-social engagement activity, then we assume that non-cooperative individuals will not engage this costly action, i.e. defectors will not be vigilant. 

Players are affected by the pressure of being watched by their neighborhood, which modifies their temptation to defect, decreasing it as the pressure (percentage of vigilant neighbors) increases. The individual temptation ${T}_{i}$ of node \emph{i} is: 

\begin{equation}
{T}_{i}=R+(T-R)(1-{m}_{i}/{k}_{i})
\end{equation}
where again ${m}_{i}$ is the number of neighbors of the node \emph{i} that are already vigilant, and ${k}_{i}$ is the degree of node \emph{i}.

The fitness of an individual is the accumulated payoff after playing ${k}_{i}$ PD games with her neighbors.

The second phase of the game is the update of individual strategies, which is performed each generation. Darwinian dynamics are introduced to promote the fittest strategy. The replicator dynamics \cite{RD1998} is the traditional approach for well-mixed populations (populations with no structure where individuals play with each other). For evolutionary models, finite populations and discrete time, the equivalent classic approach is the use of the proportional imitation rule \cite{HELBING199229, Schlag1998130}. The update of strategies is performed as follows. Let $N$ be the number of individuals in the population, $s_{i}$ the strategy the individual $i$ is playing, and $\pi_{i}$ her payoff. With the proportional imitation rule, each individual $i$ randomly choose one from her $k_{i}$ neighbors (individual $j$) and adopts her strategy with probability:

\begin{equation}
p^{t}_{ij}\equiv P \left\{s^{t}_{j} \rightarrow s^{t+1}_{i} \right\} = \left\{ \begin{array}{lcc}
(\pi^{t}_{j}-\pi^{t}_{i})/\Phi & if & \pi^{t}_{j} > \pi^{t}_{i}, \\
\\ 0 & if & \pi^{t}_{j} \leq \pi^{t}_{i}, \\
\end{array}
\right.
\end{equation}

where $\Phi = max(k_{i}, k_{j})[max(1,T)-min(0,S)]$ so that $p^{t}_{ij} \in [0,1]$.

The strategies of the individuals are updated synchronously.

\subsection{Simulations}
All simulations presented hereafter have been carried out for networks of 1000 nodes and results are averaged over at least 100 different realizations of the initial conditions. 

Players are randomly initialized as cooperators or defectors with equal probability. The cooperators are also equiprobably chosen to be either vigilant or nonvigilant, while defectors are always nonvigilant. So, on average, we start with half of the population as cooperators, half as defectors and a quarter as both cooperators and vigilant.

Statistics are measured after a transient of 100,000 generations and averaged over a time window of 100 generations, if the system has reached an stationary state defined by the slope of the average fraction of cooperators $\langle\rho\rangle$ being inferior to $10^{-2}$, if not, we let the system evolve subsequent time windows of 100 generations.

\section{Interplay between structure and dynamics}

\subsection{Cooperation dynamics over a monoplex structure} \label{sssec:mono}

\begin{figure}[h]
	\begin{center}
		\subfigure[]{%
			\label{fig:1a}
			\includegraphics[height=8.6cm, angle=-90]{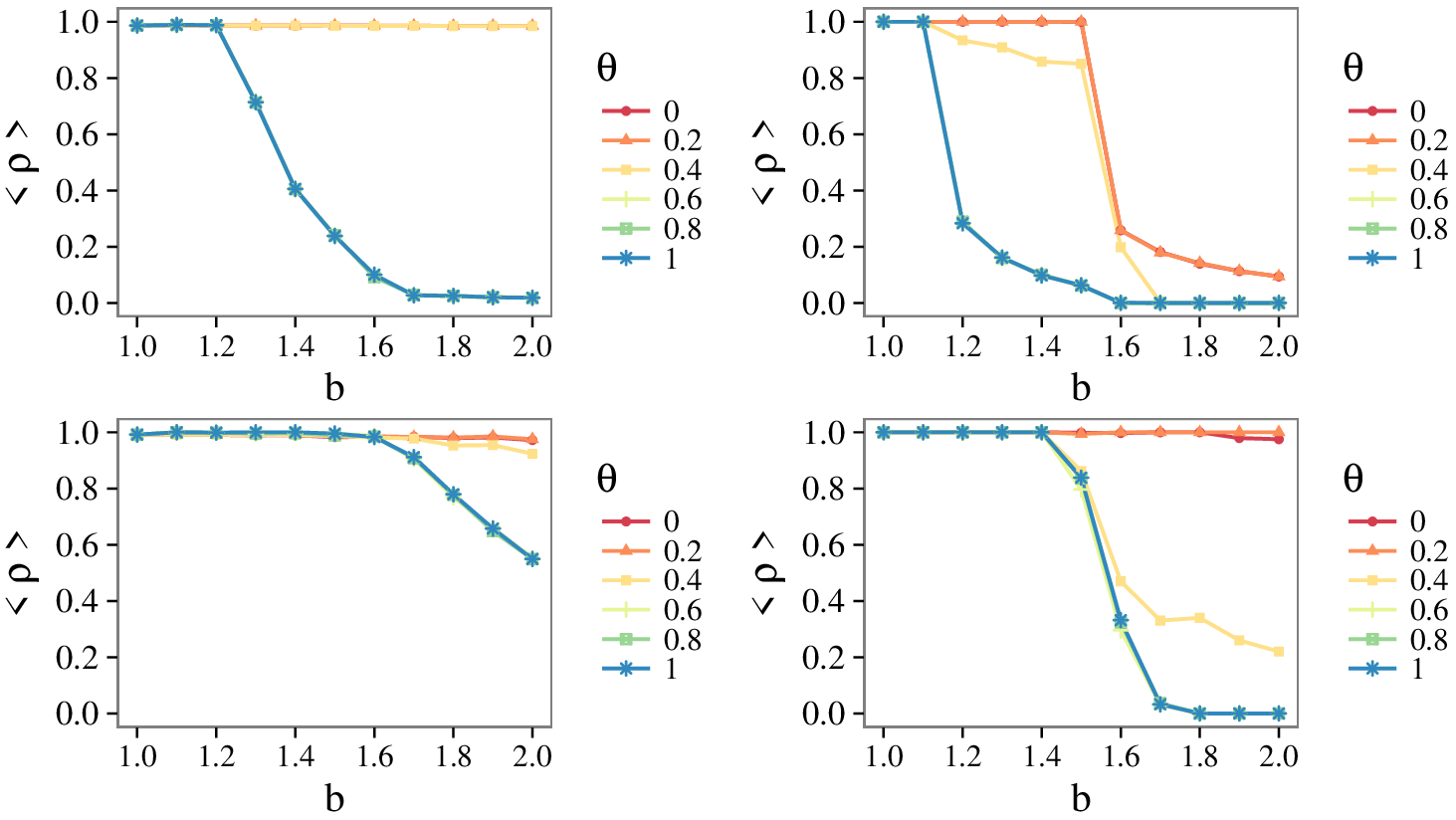}
		}%
		
		\subfigure[]{%
			\label{fig:1b}
			\includegraphics[height=8.6cm, angle=-90]{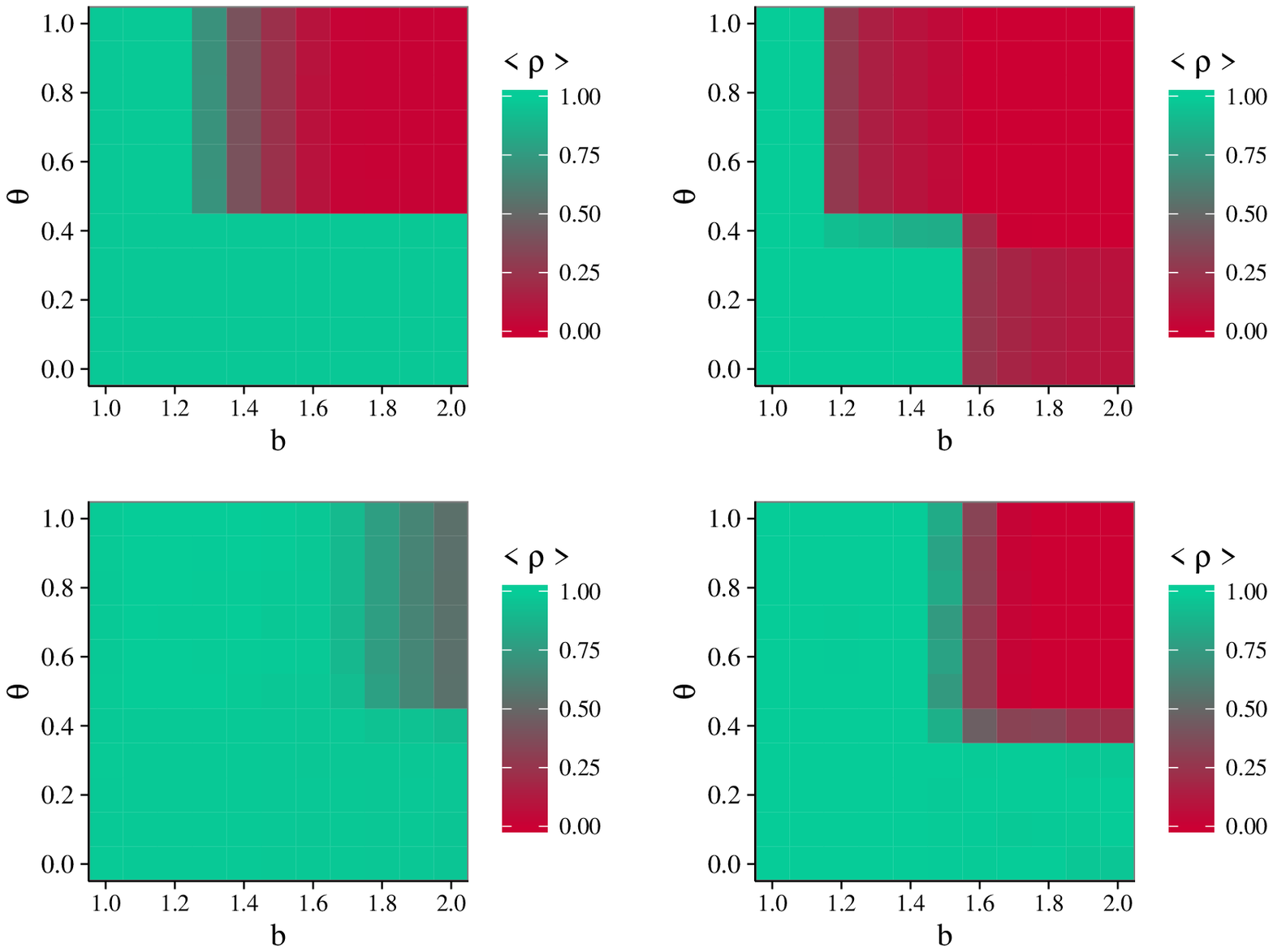}
		}%
	\end{center}
	\caption{Simulation results for an Erd\"{o}s-R\'{e}nyi network (upper panels of both subfigures) and Barab\'{a}si-Albert network (bottom panels of both subfigures) with average degree z=4 (left) and z=16 (right). Subfigure (a) show the average fraction of cooperators $\langle\rho\rangle$ as functions of the advantage of defectors $b$, and (b) show the average fraction of cooperators $\langle\rho\rangle$ as functions of the personal threshold $\theta_{i}$ of the nodes (horizontal axis) and as functions of the advantage of defectors $b$ (vertical axis). The maximum SE of all $\langle\rho\rangle$ values in the figure is 0.048.
	}%
	\label{fig:figure1}
\end{figure}

Firstly, we focus on the aforementioned two biderectionally-coupled dynamics over a monoplex network. We show that cooperation is significantly enhanced when players feel the pressure of being watched, independently of the topology and the average degree of the network (Fig.~\ref{fig:figure1}). Cooperation promotion is affected by the personal threshold $\theta_{i}$, as Fig.~\ref{fig:1a} shows.

There is a phase transition in $\langle\rho\rangle$ as functions of $\theta_{i}$. The critical point happens around 0.5 $<\theta_{i}<$ 0.8. Values of $\theta_{i}$ below this point promote cooperation for both network configurations, as Fig.~\ref{fig:1a} show, and values of $\theta_{i}$ above the critical point make no difference in the outcome of cooperation respect to the case with no vigilance network. It can be observed that values of $\theta_{i}<0.5$ are sufficient to promote cooperation and more importantly it can be quantified. The higher the threshold, the more vigilant neighbors a player need to become vigilant too, and the more difficult is to promote cooperation based on vigilance.

The average fraction of cooperators $\langle\rho\rangle$ as functions of the personal threshold $\theta_{i}$ of the nodes and as functions of the advantage of defectors $b$ is shown in Fig.~\ref{fig:1b}, where green areas represent cooperation ($\langle\rho\rangle $ greater than 0.5) and red areas represent defection ($\langle\rho\rangle$ lower than 0.5). 

The influence of the vigilance network on cooperation is far more pronounced for the Barab\'{a}si-Albert networks. In the case for z=4 (Fig.~\ref{fig:1b} bottom left panel), cooperations is fully achieved in almost all regions of the parameter space. For z=16 cooperation emerges for $b<1.5$ independently of the $\theta_{i}$ value, which does not hold for the Erd\"{o}s-R\'{e}nyi network, cooperation is fully achieved ($\langle\rho\rangle=1$) in a 70\% of the parameter space ($b<1.5$ , all $\theta_{i}$ values; and $b\geq1.5$, $\theta_{i}<0.4$).

Up to now, we have studied a population with the same personal threshold $\theta_{i}$. A more realistic approach is to have a heterogeneous population, i.e. to initialize the population with random $\theta_{i}$ values. If we initialize the $\theta_{i}$ of the population with values drawn from a uniform distribution U[0,1], the expected $<\theta_{i}> = 0.5$ and indeed the results for the average fraction of cooperators $\langle\rho\rangle$ are undistinguishable from the ones obtained when  $\theta_{i}= 0.5$. This also holds for the duplex network case in Fig.~\ref{fig:duplex1}.


\subsection{Influence of cost of the vigilance action} \label{sssec:cost}

\begin{figure}[h]
	\begin{center}
		\includegraphics[width=8.6cm, angle=0]{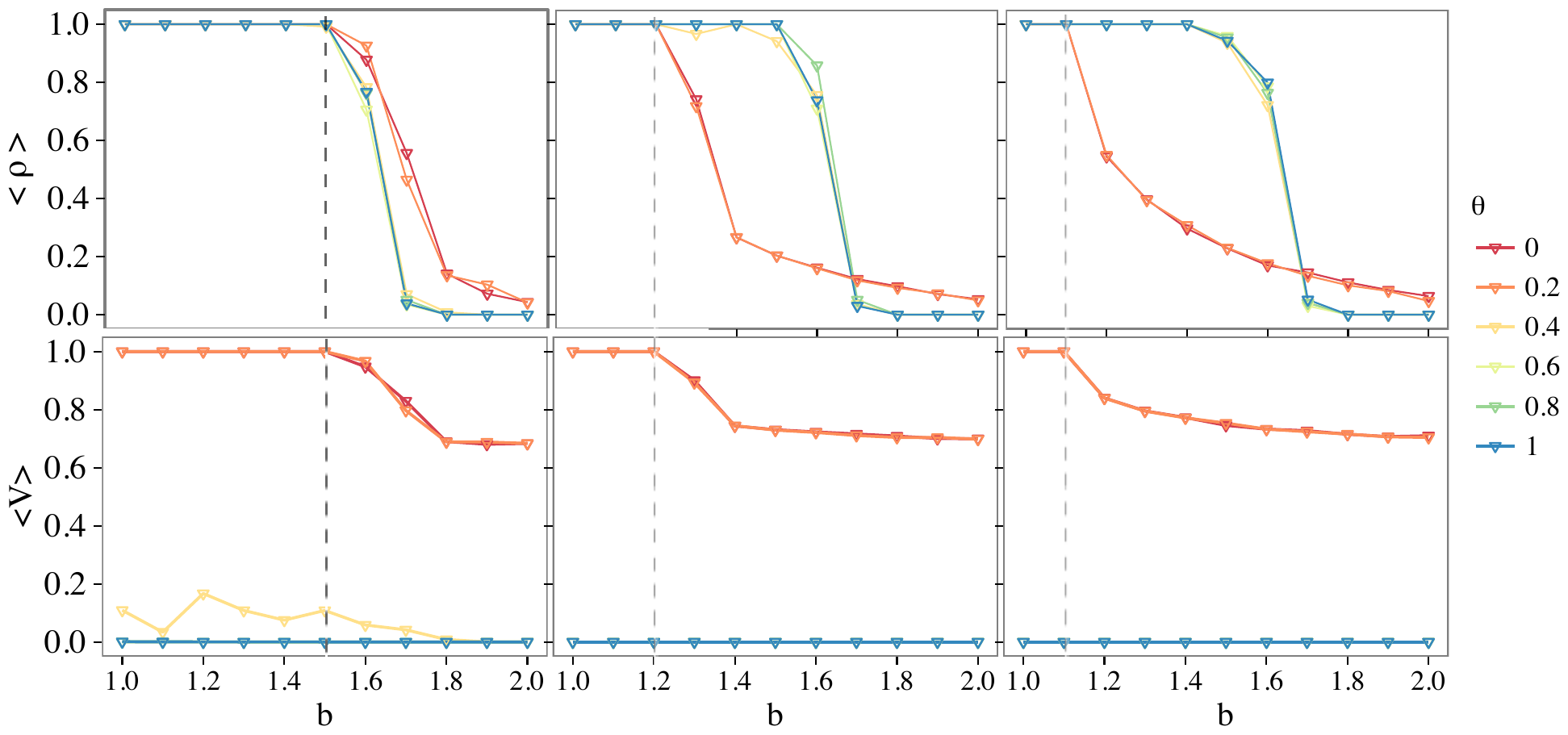}	
	\end{center}
	\caption{Average fraction of cooperators $\langle\rho\rangle$ (top panels) and average fraction of vigilant players $\langle V \rangle$ (down panels) for Barab\'{a}si-Albert networks with z=16 and cost of vigilance $Cv$ 0.25R (left), 0.5R (middle), 0.75R (right). The higher the cost of vigilance, the lower the average fraction of cooperators $\langle\rho\rangle$ in the regions of $b$ where the population is not fully vigilant. The maximum SE of all $\langle\rho\rangle$ values in the figure is 0.023.}
	\label{fig:figure2}
\end{figure}

\begin{figure}[h]
	\begin{center}
		\subfigure[Vigilance networks as Erd\"{o}s-R\'{e}nyi. Game networks as Erd\"{o}s-R\'{e}nyi (top) and Barab\'{a}si-Albert (bottom). Both layers with the same degree, z=4 (left) or z=16 (right).]{%
			\label{fig:duplex1a}
			\includegraphics[height=8.6cm, angle=-90]{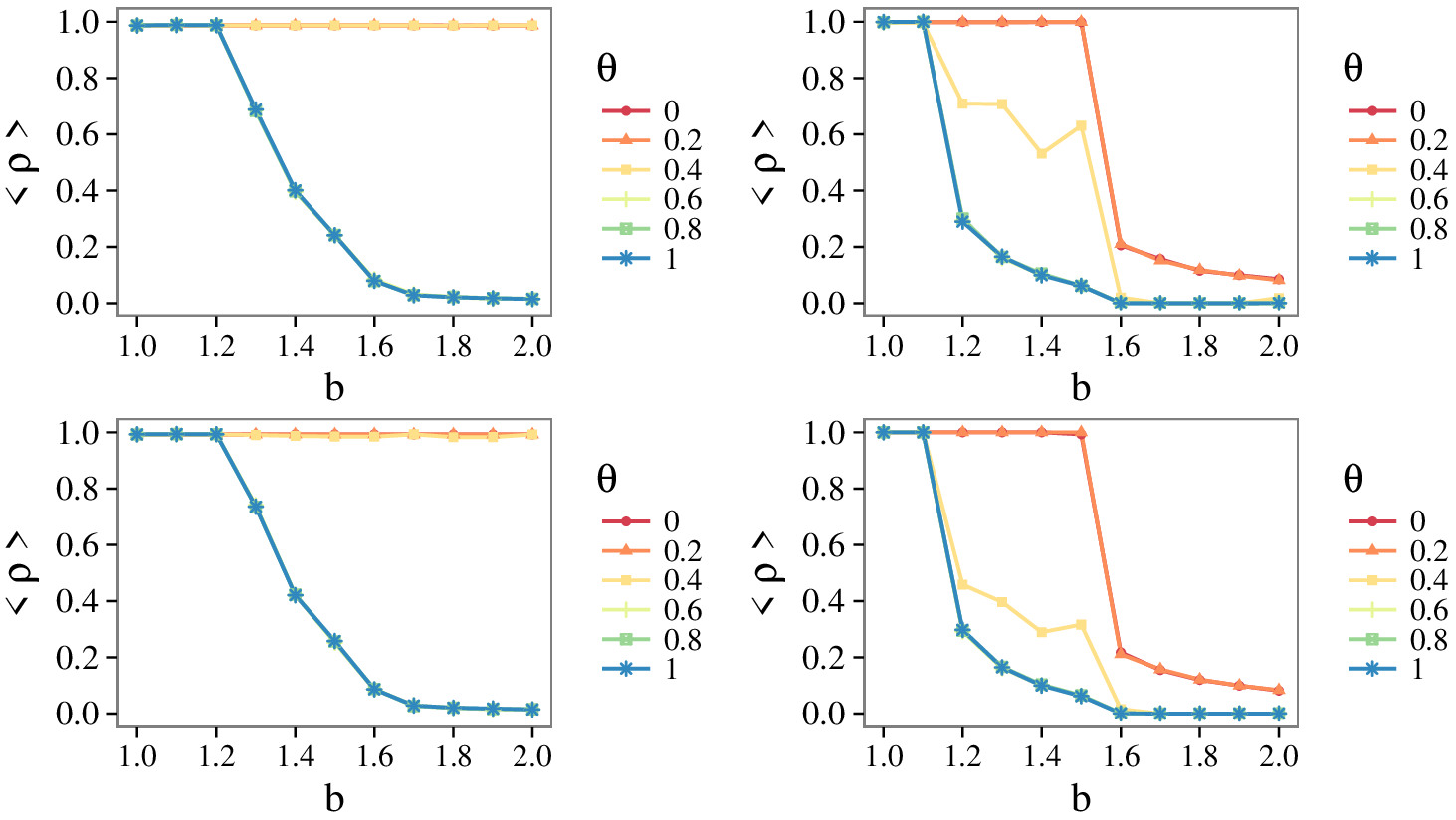}
		}%
		
		\subfigure[Vigilance networks as Barab\'{a}si-Albert. Game networks as Erd\"{o}s-R\'{e}nyi (top) and Barab\'{a}si-Albert (bottom). Both layers with the same degree, z=4 (left) or z=16 (right).]{%
			\label{fig:duplex1b}
			\includegraphics[height=8.6cm, angle=-90]{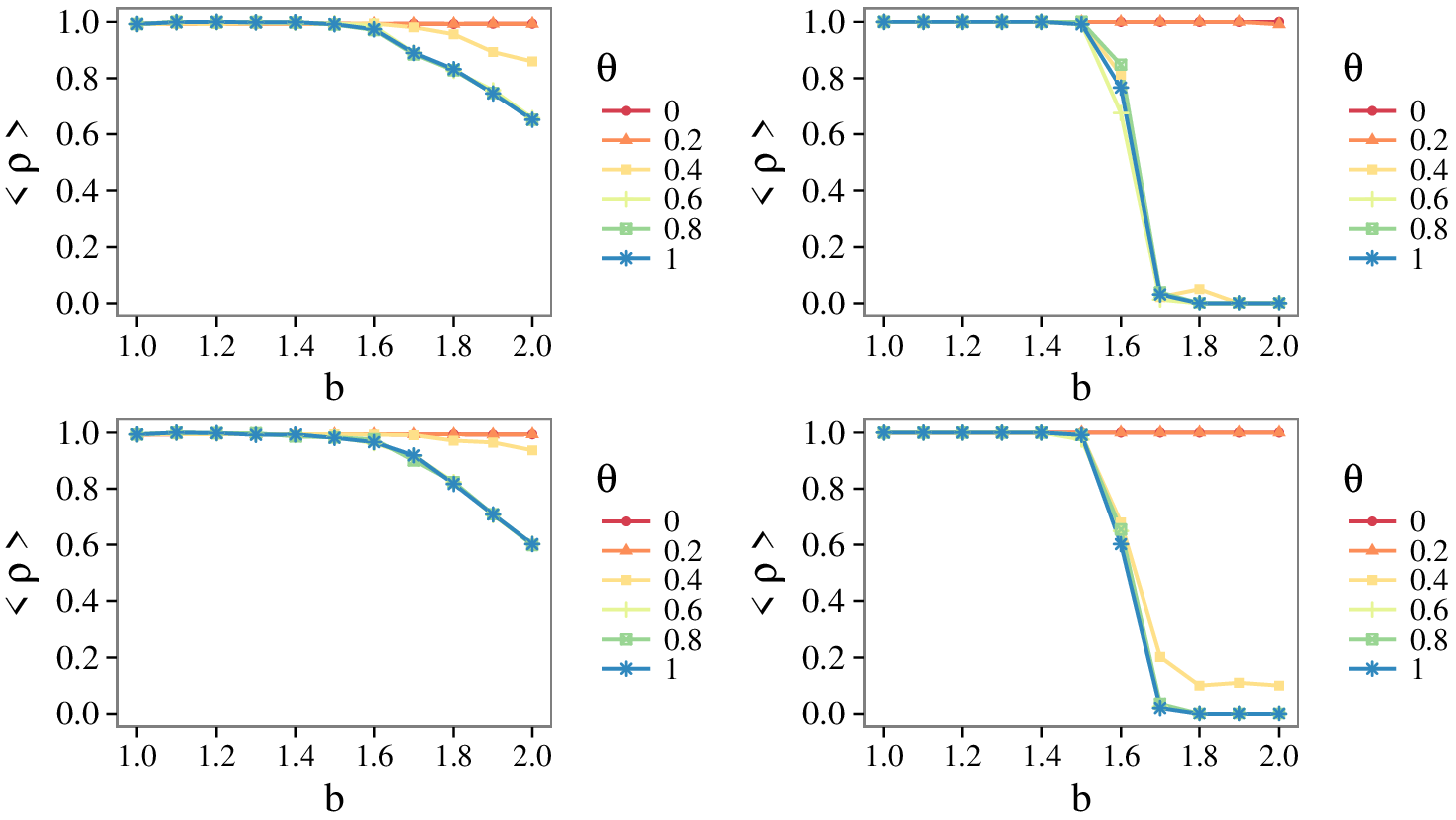}
		}%
	\end{center}
	\caption{Average fraction of cooperators $\langle\rho\rangle$ for a duplex structure of networks. The combination of network topologies (Erd\"{o}s-R\'{e}nyi and Barab\'{a}si-Albert networks) and average degrees (both layers with the same degree, z=4 or z=16) are explored.  The vigilance network drives the cooperative outcome of the game dynamics. The vigilance network promotes cooperation as in the monoplex scenario (Fig.~\ref{fig:figure1}). The maximum SE of all $\langle\rho\rangle$ values in the figure is 0.05.}
	\label{fig:duplex1}
\end{figure}

So far our analysis of the interplay between vigilance and cooperation assumed no cost for the vigilance action. However, a more realistic hypothesis is that vigilance comes at a certain cost for the action. We have introduced this cost in the following way: we consider that every agent has to afford a vigilance cost each generation, which is a fraction ($Cv$) of her reward in the game $R$. Fig.~\ref{fig:figure2} shows the influence of $Cv$ in the outcome of cooperation for the Barab\'{a}si-Albert network. 

When the cost of vigilance is high ($0.5R$ and $0.75R$, middle and left panels of Fig.~\ref{fig:figure2}), the vigilance network can promote cooperation and defection depending on the value of the temptation parameter $b$. Note that the percentage of vigilant nodes in the population $\langle V \rangle$ corresponds to the critical point in $b$ (not the same for all $\theta_{i}$) from which full cooperation is dismantled. This critical point in $b$ where $\langle V \rangle$ starts to be inferior to 1 can be used as a predictor for the critical point in $b$ above which full cooperation disappears, as the dashed vertical grey lines in Fig.~\ref{fig:figure2} exemplify for several arbitrary $\theta_{i}$ values. As the cost of vigilance $Cv$ increases, the critical point (above which full cooperation is dismantled) is shifted to lower values of $b$.

\subsection{Cooperation dynamics over a duplex structure} \label{sssec:duplex}

We extend the analysis to a duplex structure of networks: a network for the game dynamics and a network for the vigilance dynamics. We study the bidirectional coupling of both layers. 

Firstly, we will consider that both layers have the same average degree and we will study the influence of network topology. For a vigilance network with a given average degree and topology, we study the influence of vigilance layer for different game networks topologies. For example, for a vigilance Barab\'{a}si-Albert network with z=16 (Fig.~\ref{fig:duplex1b} right panels), the average fraction of cooperators $\langle\rho\rangle$ is shown for the case where the game network is a Barab\'{a}si-Albert network (upper panel) and an Erd\"{o}s-R\'{e}nyi network (bottom panel). For the four vigilance network configurations studied, the vigilance network has a significant influence in the dynamics of cooperation so that there is no significant difference in the stationary average fraction of cooperators $\langle\rho\rangle$ for different game network topologies, as it can be seen at Fig.~\ref{fig:duplex1}. The outcome of cooperation is lead by the vigilance network. 

\begin{figure}
	\begin{center}
		\includegraphics[height=8.6cm, angle=-90]{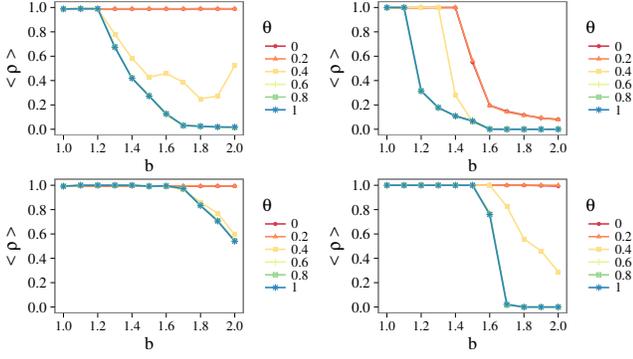}	
	\end{center}
	\caption{Average fraction of cooperators $\langle\rho\rangle$ for a duplex structure of networks, where layers have different average degrees. Vigilance and game networks both as Erd\"{o}s-R\'{e}nyi (top panels),  vigilance and game networks both as Barab\'{a}si-Albert (bottom panels). Vigilance networks with z=4 and game networks with z= 16 (left panels), vigilance networks with z=16 and game networks with z= 4 (right panels). The maximum SE of all $\langle\rho\rangle$ values in the figure is 0.05.}
	\label{fig:duplex2}
\end{figure}

Now, let's focus on the case where vigilance and game layers have not equal average degree. As Fig.~\ref{fig:duplex2} shows, the average degree of the vigilance layer dominates the dynamics of cooperation. Indeed, there is no significant difference in the average fraction of cooperators $\langle\rho\rangle$ obtained where the game layer has the same average degree of the vigilance network or it is different, and also when the topology of the game layer is different. It can be seen that each panel in Fig.~\ref{fig:duplex2} shows the same average fraction of cooperators $\langle\rho\rangle$ that the one in Fig.~\ref{fig:duplex1}. For example, top-left panel in Fig.~\ref{fig:duplex2} corresponds to Fig.~\ref{fig:duplex1a} left panels, and it also holds for the other panels in Fig.~\ref{fig:duplex2} and its correspondence in Fig.~\ref{fig:duplex1}. Vigilance networks with higher average degree hinder cooperation, since it is more difficult for an agent to fulfill her threshold to become vigilant, and therefore, the diffusion of the vigilance dynamics is more costly.

Now we study the impact of correlated multiplexity, i. e. layer-degree correlations, since in real-world complex systems the degree of nodes in the different layers of the multiplex structure are not randomly distributed but correlated. We focus this study on a duplex structure where both layers are Barab\'{a}si-Albert networks, as the majority of real-world social networks present scale free degree distributions with exponent between 2 and 3.

\begin{figure}
	\addtolength{\subfigcapskip}{1.5mm} 
	\begin{center}
		\subfigure[ Maximally positive correlated multiplexity.]{%
			\label{fig:duplex3a}
			\includegraphics[height=8.6cm, angle=-90]{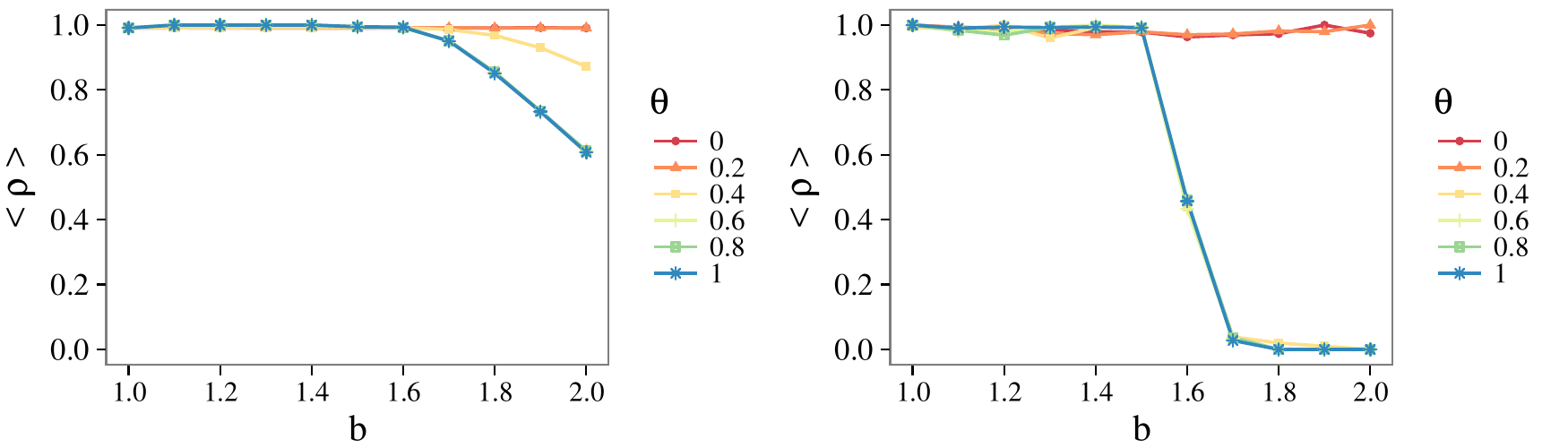}
		}%
		
		\subfigure[ Maximally negative correlated multiplexity.]{%
			\label{fig:duplex3b}
			\includegraphics[height=8.6cm, angle=-90]{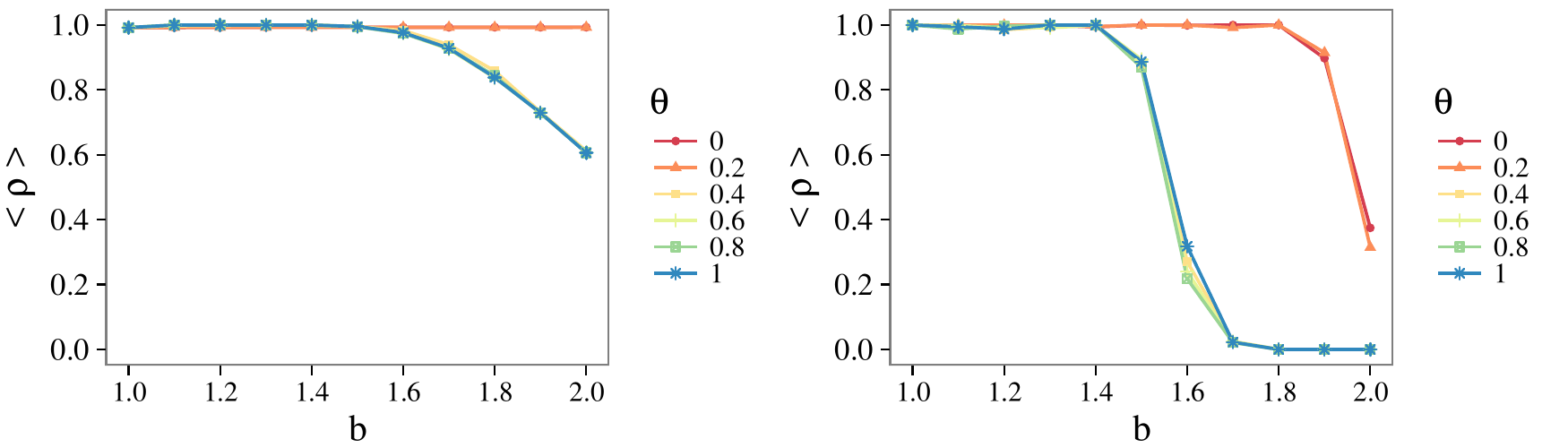}
		}%
	\end{center}
	\caption{Average fraction of cooperators $\langle\rho\rangle$ for a duplex structure of Barab\'{a}si-Albert networks. The degree distributions of the layers are maximally positive correlated (a) and maximally negative correlated (b). The average degree of both layers of the duplex is the same, and takes the values z=4 (left) and z=16 (right).  The maximum SE of all $\langle\rho\rangle$ values in the figure is 0.044.}
	\label{fig:duplex3}
\end{figure}

Results for layers with average degree distribution $z=4$ (Fig.~\ref{fig:duplex3} left panels) do not show to be influenced significantly by the correlated multiplexity of their layers. It is not the same for higher average degrees (Fig.~\ref{fig:duplex3} right panels). When the degree distribution of game and vigilance layers are maximally positive correlated (Fig.~\ref{fig:duplex3a}), cooperation is fairly promoted ($\theta_{i}\leq 0.3$ end up with $\langle\rho\rangle$ = 1), compared to the uncorrelated scenario (Fig.~\ref{fig:1a} bottom panels). In the opposite case, where layers are maximally negative correlated (Fig.~\ref{fig:duplex3b}), cooperation is drastically hindered. The critical point where cooperation is dismantled is shifted to lower $b$ values for all $\theta$ values. In fact, for values of $b$ larger than 1.8 full cooperation in not achieved for any $\theta$ value.

\subsection{Vigilance dynamics with giving up option}  \label{sssec:giveup}

So far the dynamics for the vigilance layer accounted that once an agent has become vigilant, she cannot scape this situation. In real situations, people can stop feeling social pressure or just decide to change their opinion/action. In this subsection, we take into account the possibility of giving up being vigilant. We approach this by two means: (1) using the threshold of vigilance in a reverse way, and (2) with a probability of giving up vigilant. We analyze this situation for a duplex of  Barab\'{a}si-Albert networks.

Firstly, lets consider the vigilance dynamics as composed by two processes: becoming vigilant and becoming nonvigilant. The process of becoming vigilant is the same as in eq.~\ref{eq:vigilance}. The process of giving up vigilance ${V}^{1 \rightarrow 0}_{i}$ follows an inverse threshold model, where an agent become nonvigilant is there is not enough vigilant actors in her neighborhood: 
\begin{equation}\label{eq:vigilanceback}
{V}^{1 \rightarrow 0}_{i}({m}_{i},{k}_{i})= \left\{ \begin{array}{lcc}
             1 & if & {m}_{i}/{k}_{i} \leq {\theta}_{i}, \\
             \\ 0 & if &   {m}_{i}/{k}_{i} > {\theta}_{i}, \\
             \end{array}
   \right.
\end{equation}

\begin{figure}
	\addtolength{\subfigcapskip}{1.5mm} 
	\begin{center}
		\subfigure[ Inverse threshold model.]{%
			\label{fig:6a}
			\includegraphics[height=8.6cm, angle=-90]{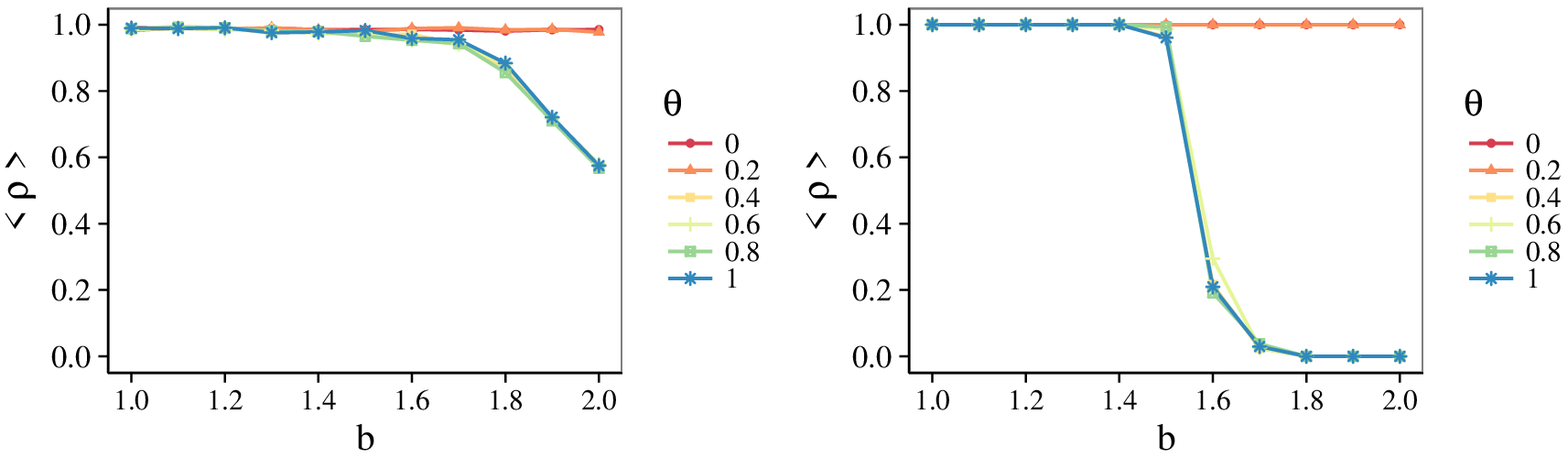}
		}%
		
		\subfigure[ Probability of giving up vigilance.]{%
			\label{fig:6b}
			\includegraphics[height=8.6cm, angle=-90]{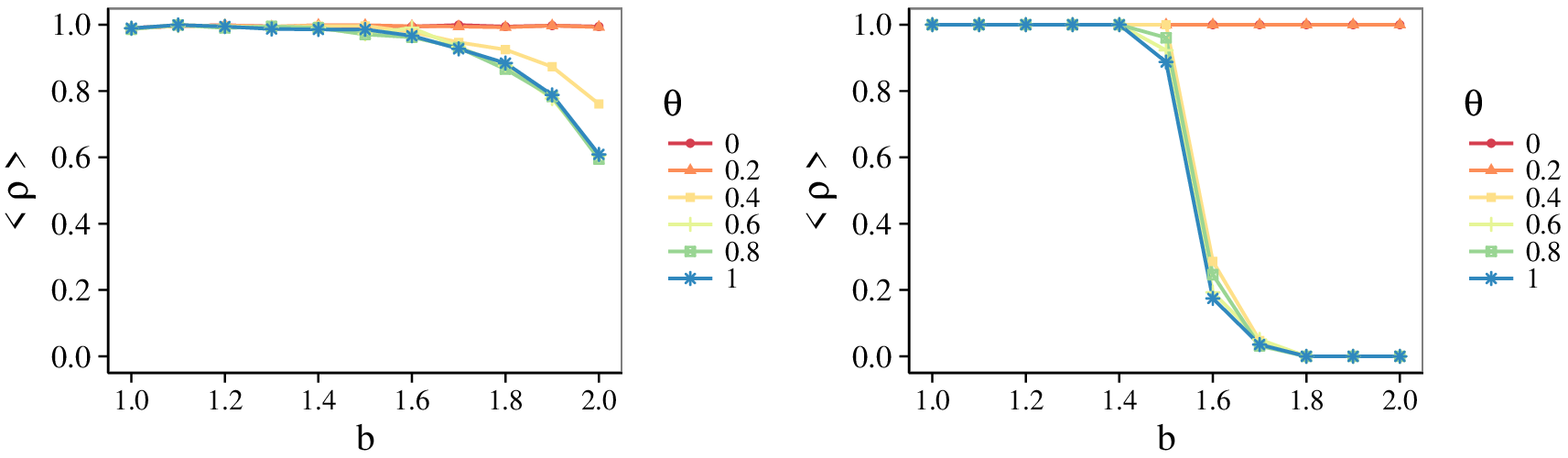}
		}%
	\end{center}
	\caption{Average fraction of cooperators $\langle\rho\rangle$ for a duplex structure of Barab\'{a}si-Albert networks when actors can give up being vigilant, (a) is there is no enough social pressure in the neighborhood, or (b) with a probability p=0.05. The average degree of both layers of the duplex is the same, and takes the values z=4 (left) and z=16 (right).  The maximum SE of all $\langle\rho\rangle$ values in the figure is 0.03.}
	\label{fig:6}
\end{figure}

As Fig.~\ref{fig:6a} shows, the average fraction of cooperation $\langle\rho\rangle$ is still influenced by the threshold of vigilance $\theta_{i}$. Cooperation is most costly now, and is only promoted when actors do not need much social pressure to become vigilant ($\theta_{i}<0.4$ which is inferior to the threshold needed to promote cooperation in the case in Fig.~\ref{fig:duplex1b} (down panels).

If actors become vigilant again following a threshold model (eq.~\ref{eq:vigilance}) but can give up vigilance with a probability $p=0.05$, we find comparable results (Fig.~\ref{fig:6b}). Slightly higher cooperation levels can be found for the case with z=4 and $\theta_{i}=0.4$ (Fig.~\ref{fig:6b}, left panel) related to the scenario in Fig.~\ref{fig:6a}. Broadly, we can conclude that giving the option of stopping the vigilance action do not hinders cooperation, but there is few $\theta_{i}$ values for which cooperation is enhanced.

\section{Conclusions}
Summarizing, we have presented a computational analysis of the interplay between vigilance network dynamics and game dynamics, showing that the pressure of being watched plays a significant role in the outcome of cooperation, having both the effect of booster and dismantler. We show that the conditions for the observability effect to emerge are not trivial, and subsequent experimentation with real humans would be of much interest to fully understand the impact of being watched on cooperation.
These results clear the ground for a framework to quantify the promotion of cooperation in structured populations that use vigilance to enhance pro-social behavior.

\begin{acknowledgments}
The author acknowledge support from the project H2020 FET OPEN RIA IBSEN/662725, and from Institute of Physics of Cantabria (IFCA-CSIC) for providing access to Altamira supercomputer.
\end{acknowledgments}


\nocite{*}

\bibliography{myrefs}

\end{document}